\documentclass[aps,superscriptaddress,showpacs,amsmath,amssymb,twocolumn,nofootinbib]{revtex4-1}

\usepackage[dvips]{graphicx}
\usepackage{float}
\usepackage[export]{adjustbox} 
\usepackage{dcolumn}
\usepackage{bm}
\usepackage{braket}
\usepackage{amsthm}
\usepackage{caption}
\usepackage{subcaption}
\usepackage{bbold}
\usepackage{amsmath}
\usepackage{amsfonts}
\usepackage{array}
\usepackage{ulem}
\usepackage[thinlines]{easytable}
\usepackage{color}

\makeindex
\begin{document}

\title{The evolution of magnetic dipole strength in $^{100-140}$ Sn isotope chain and quenching of nucleon g factors}
\author{G. Kru{\v z}i{\' c}} \email{goran.kruzic@ericsson.com}
 \affiliation{Department of Physics, Faculty of Science, University of Zagreb, Bijeni{\v c}ka c. 32, HR-10000, Zagreb, Croatia}
 \affiliation{Research department, Ericsson - Nikola Tesla, Krapinska 45, HR - 10000, Zagreb, Croatia}
\author{T. Oishi}
 \email{toishi@phy.hr}
 \affiliation{Department of Physics, Faculty of Science, University of Zagreb, Bijeni{\v c}ka c. 32, HR-10000, Zagreb, Croatia}
\author{N. Paar}%
 \email{npaar@phy.hr}
 \affiliation{Department of Physics, Faculty of Science, University of Zagreb, Bijeni{\v c}ka c. 32, HR-10000, Zagreb, Croatia}
%
%

\begin{abstract}
{\noindent
The evolution of electromagnetic transitions along isotope chains is of particular importance
for the nuclear structure and dynamics, as well as for the r-process nucleosynthesis. Recent
measurement of inelastic proton scattering on even-even $^{112-124}$Sn isotopes provides a novel
insight into the isotopic dependence of E1 and M1 strength distributions. We investigate 
M1 transitions in even-even $^{100-140}$Sn isotopes from a theoretical perspective, based on relativistic 
nuclear energy density functional.
The M1 transition strength distribution is characterized by an interplay between single and double-peak structures, 
that can be understood from the evolution of single-particle states, their occupations governed by the
pairing correlations, and two-quasiparticle transitions involved. It is shown that discrepancy between model calculations and experiments for the M1 transition strength is considerably reduced than previously known, and the quenching of the $g$ factors for the free nucleons needed to reproduce the experimental data on M1 transition strength 
amounts $g_{eff}/g_{free}$=0.80-0.93. Since some of the M1 strength above the neutron threshold may be missing in the inelastic proton scattering measurement, further experimental studies are required to
confirm if only small modifications of the bare $g$ factors are actually needed when applied in finite nuclei.
}
\end{abstract}
\maketitle

\section{\label{sec:INTRO} Introduction}

Electromagnetic transitions play an essential role in single-particle and collective nuclear excitations, important not only for the properties of finite nuclei, but also for astrophysically relevant processes.
Various aspects of magnetic dipole (M1) excitations have been considered both in experimental 
and theoretical studies \cite{Richter02, 1995Richter, Richter01, Kneissl01, Pietralla01, 2015Goriely, 2016Goriely} due to their 
relevance for diverse nuclear properties associated e.g., to unnatural-parity states, spin-orbit splittings and tensor force effects.
Specifically, M1 spin-flip excitations are analog of Gamow-Teller (GT) transitions, meaning that, at the operator level, 
the dominant M1 isovector component is the synonym to the zeroth component of the GT process, 
and can serve as probe for calculations of inelastic neutrino-nucleus cross section \cite{Langanke01,Langanke02}. 
This process is hard to measure but essential in supernova physics, and also relevant for the r-process 
nucleosynthesis calculations~\cite{Langanke2008,Loens2012,2015Goriely,2016Goriely}. 
The isovector spin-flip M1 response is also relevant for applications related to the design of
nuclear reactors~\cite{Chadwick2011}, for the understanding of single-particle properties, spin-orbit interaction, and shell closures from stable nuclei 
toward limits of stability~\cite{Otsuka2005,Otsuka2010,Nesterenko01,Nesterenko02,Tselyaev01}. It also relates to resolving the problem of quenching of the spin-isopin response 
in nuclei that is necessary for reliable description of double beta decay 
matrix elements~\cite{Vergados2012}. 
In deformed nuclei, another type of M1 excitations has been extensively studied, 
known as scissors mode, where the orbital part of M1 operator plays a dominant role in a way that protons and neutrons oscillate with opposite phase around the core \cite{Arteaga01, Arteaga02, Richter01, Richter03, Otsuka01, Bohle01, Bohle02, Schwengner01, Balbutsev01, Repko01}. 
In any nuclei undergoing experimental investigation, 
there are simultaneously present $E\lambda$ and $M\lambda$ multipole excitations, 
where the electric dipole (E1) and electric quadrupole (E2) responses \cite{2019OP, Dang01, Yoshida01, Terasaki01, Ebata01, Stetcu01} dominate over the M1 response \cite{Tamii01, Fujita, 2016Birkhan, Laszewski, Laszewski02, Laszewski03, Laszewski04, Raman01, Raman02}. 
Thus, it is a rather challenging task to extract M1-related observables in a whole energy range.
Even for the nuclides accessible by experiments, their full information 
on the M1 response has not been complete.

The M1 transitions have been studied in various theoretical approaches; for more details see review~\cite{Richter01} and references therein. 
In particular, the properties of the M1 resonances have been investigated in the framework based on the Skryme functionals \cite{Nesterenko01, Nesterenko02, Tselyaev01}. It has been shown that the results for the spin-flip resonance obtained by using different parameterizations do not appear as convincing interpretation of the experimental results.
Additional effects have been explored in order to resolve this issue, e.g., the isovector-M1 response 
versus isospin-mixed responses, and the role of tensor and vector spin-orbit interactions~\cite{Nesterenko01, Nesterenko02}. 
In a recent analysis~\cite{Speth_2020}, it has been shown that while modern Skyrme functionals successfully reproduce electric excitations, there are difficulties to describe magnetic transitions, and further developments in the the spin channel are called for~\cite{Speth_2020}.
The properties of M1 excitations have also recently been studied in the framework of relativistic
nuclear energy density functional \cite{2020Oishi, 2020Kruzic}, and the role of the residual interaction including pseudovector channel has been explored.
Also, the pairing correlation has been found as another, important ingredient to control the M1-excitation properties \cite{2019OP, 2020Oishi}.
Previous systematic study of M1 strength functions based on D1M Gogny force showed
that  the available experimental data could be reproduced if the calculated strength is shifted
globally by about 2 MeV and increased by an empirical factor of 2 \cite{2016Goriely}.
Therefore, for the complete understanding of the M1 excitation properties, further systematic analysis is 
necessary. 

Recently, an experimental study based on the inelastic proton scattering provided novel data
on E1 and M1 strength distributions along the even-even $^{112-124}$Sn isotope chain \cite{Bassauer2020}.
The resulting photoabsorption cross sections derived from the E1 and M1 strength distributions showed 
significant differences when compared to those from previous $(\gamma,xn)$ experiments \cite{1989Alarcon, 1998Govaert}.
The aim of this work is to explore the properties on M1 excitations in a broad range of even-even Sn nuclides, and
examine the model calculations in view of the recent experimental data from Ref. \cite{Bassauer2020}.
These data could allow us to establish an essential link between the M1 observables and theoretical models and improve our
understanding of the role of M1 transitions in modeling radiative neutron capture cross sections of relevance for nucleosynthesis.

Numerous studies in the past addressed possible quenching effects in the spin $g$ factors
when applied in finite nuclei \cite{Richter03, Nesterenko01, Nesterenko02, Dehesa}. The quenching factors
can be obtained by normalizing the calculations on M1 transitions to the experimental data. 
Accordingly, the free value of the $g$ factor ($g_{free}$) is often considerably quenched, leading to its effective value that was previously reported mainly as $g_{eff} \approx$ 0.6-0.75 $g_{free}$ \cite{Heyde2010,1989Alarcon,2009Vesely,2010Nest,1998VNC,Ichimura2006}.
Therefore, in view of the quenching of the $g$ factors in finite nuclei it is interesting to explore how 
the novel inelastic proton scattering data \cite{Bassauer2020} compare to the results on M1 transitions in the
framework of the relativistic energy density functional. As previously discussed, one of the most important 
mechanisms responsible for the quenching is mixing with higher order 
configurations \cite{Heyde2010,Bertsch1982,Ichimura2006,1988Takayanagi,1989Kamerd,1993Kamerd}. 
Additional effects have been suggested to arise from the core excitation \cite{Brown1983}, and 
the meson-exchange current effect \cite{1977Dehesa,Marucci2008}.
Since these effects
are not included in this work, it is interesting
to explore to what extent the recent data from inelastic proton scattering can be reproduced 
by using the microscopic theory framework at the QRPA level,
based on the advanced density-dependent relativistic point coupling interactions and supplemented 
with the pairing correlations in a consistent approach.

For the evaluation of nuclear $g$ factors in medium, we briefly mention the other probe than M1, namely, the E2 transition.
The E2 transition can also provide the information on the quenching of $g$ factors in the nuclear excited states \cite{Terasaki2002, Almond2015}.
Even though it is a fascinating topic, in the present study, we skip the further discussion on the E2 transition, and concentrate on the M1 properties.

The paper is organized as follows.
In Sec. \ref{sec:FORMLSM}, theoretical framework of the present work is summarized, including the relativistic point-coupling interaction, relativistic quasiparticle random-phase approximation (QRPA), as well as magnetic-dipole transitions.
Section \ref{sec:Sn_ISOTOPES} is devoted to the results of the model calculations of M1 transitions in the
Sn isotope chain. Finally, in Sec. \ref{sec:SUM}, we summarize this work.

\section{\label{sec:FORMLSM} Theory framework}
The M1 transitions
from $0^{+}$ ground state (GS) to $1^{+}$ excited states are studied in the framework of
relativistic nuclear energy density functional (RNEDF), assuming the spherical symmetry. 
Various implementations of the RNEDF have been successful in the description of
nuclear excitation properties \cite{Paar2005,Paar2009,Niu2013,Niu2009,Khan2011,Yuksel2020}, astrophysically relevant weak interaction processes \cite{Paar2008,Samana2011,Fantina2012,Vale2016,Petkovic2019}, and nuclear equation of state \cite{RocaMaza2018,Mondal2016,Paar2014}.
Here we give a brief overview of the theory framework extended for the study of M1 transitions;
for more details also see Refs. \cite{2020Oishi, 2020Kruzic}.
The nuclear ground state is described in the relativistic Hartree-Bogoliubov
(RHB) model using the relativistic point-coupling interaction with density-dependent
couplings \cite{Niksic01, Niksic03,Yuksel2019}. The formalism is established through
the effective Lagrangian density, that includes the nucleon's four-point interactions in the isoscalar-scalar (IS-S), 
isoscalar-vector (IS-V), and isovector-vector (IV-V) channels. In this work the DD-PC1 parameterization is
used in model calculations \cite{Niksic01}. The pairing correlations in the RHB model are described by the
 pairing part of the phenomenological Gogny interaction \cite{Vretenar_01}, with the D1S parameterization \cite{Berger}.
The RHB calculations in this work are performed in the computational framework given in Ref. \cite{Niksic01}. 

For the description of M1 transitions up to the one-body-operator level, 
we employ the relativistic quasiparticle random phase approximation (RQRPA) based on the RHB ground 
state \cite{Paar2003,2020Oishi, 2020Kruzic}.
In the limit of small amplitude oscillations, the RQRPA matrix equations read, 
\begin{equation}
\begin{pmatrix}
    A^J      &  B^J \\
    B^{*J}   &  A^{*J}  
\end{pmatrix} 
\begin{pmatrix}
    X^{\nu,JM} \\
    Y^{\nu, JM}   
\end{pmatrix} 
=
\hbar\omega_{\nu}
\begin{pmatrix}
    1   &  0 \\
    0   &  -1  
\end{pmatrix} 
\begin{pmatrix}
    X^{\nu,JM} \\
    Y^{\nu, JM}   
\end{pmatrix} 
\end{equation}
with $\hbar\omega_{\nu} = E_{\nu} - E_{0}$, where $E_{0}$ and $E_{\nu}$ are the RHB-ground and excited state energies of the many particle system, respectively. 

In order to describe the unnatural parity transitions such as M1, in the residual RQRPA interaction
we also include the isovector-pseudovector (IV-PV) interaction, given by the Lagrangian density
\begin {equation}
\begin{aligned}
\mathcal{L}_{\rm IV-PV} = -\frac{1}{2}\alpha_{\rm IV-PV} \lbrack  \bar{\Psi}_{N} \gamma^{5} \gamma^{\mu} \vec{\tau}  \Psi_{N} \rbrack \cdot \lbrack  \bar{\Psi}_{N} \gamma^{5}  \gamma_{\mu} \vec{\tau} \Psi_{N}  \rbrack.
\end{aligned}
\end {equation}
The strength parameter, $\alpha_{\rm IV-PV} = 0.53\ \rm MeV fm^{3}$, is considered 
as a parameter, which is constrained by the experimental data on the 
M1 transitions of selected nuclei, as it is given in our previous work \cite{2020Kruzic}. 
Note that the IV-PV term does not contribute in the RHB calculation of 
the ground state because it would lead to parity violating mean field, and thus, its strength parameter cannot be constrained by the bulk properties of nuclei.

The solution of R(Q)RPA equations provides discrete spectrum of the $\nu$th excited state, $B_{\rm M1}(\omega_{\nu})$, for the magnetic-dipole operator $\hat{\mu}_{\rm M1}$ \cite{Paar2003}. That is,
\begin{equation}
\label{BM1expr}
\begin{aligned}
B_{\rm M1}(\omega_{\nu}) = &\Big\vert \sum_{\kappa \kappa'} \Big( X^{\nu,J0}_{\kappa \kappa'} \langle \kappa||\hat{\mu}_{\rm M1} ||\kappa' \rangle
\\&
+
(-1)^{j_{\kappa} - j_{\kappa'} + J}Y^{\nu,J0}_{\kappa \kappa'} \langle \kappa'||\hat{\mu}_{\rm M1} ||\kappa \rangle
      \Big) 
\\
&\times \Big(u_{\kappa}v_{\kappa'} + (-1)^{J}v_{\kappa}u_{\kappa'} \Big)
\Big\vert^{2}
\end{aligned}
\end{equation}
with $J=1$, where $\kappa$ and $\kappa'$ are quantum numbers which are denoting single-particle states in the canonical basis \cite{Paar2003}. 
The $u_{\kappa}$ and $v_{\kappa}$ are the RHB occupation coefficients. 
For demonstration purpose, this quantity is convoluted with the Lorentzian distribution~\cite{Paar2003},
\begin{equation}\label{eq:5}
\begin{aligned}
R_{\rm M1}(E) = \sum_{\nu} B_{\rm M1}(\omega_{\nu}) \frac{1}{\pi} \frac{\Gamma/2}{ (E - \hbar\omega_{\nu})^{2} + (\Gamma/2)^{2}  },
\end{aligned}
\end{equation}
where the Lorentzian width is set as $\Gamma = 1.0 \rm \ MeV$. In the RQRPA calculations we use the $g$-factors of the bare nucleons \cite{Ring01}.

The present formalism assumes a spherical symmetry to be applied for Sn isotopes. 
On the other side, in the recent shell-model study~\cite{Togashi2018},
the possibility of ground-state deformation especially in $^{102-116}$Sn and the corresponding shell evolution 
have been investigated. 
In order to account for these effects in the RNEDF framework, however, further developments are needed. 
With attention on this point, in the following sections, we show the results for the chain of Sn isotopes, in order to highlight the systematic aspects.

\begin{figure}[t]
\includegraphics[scale=0.9]{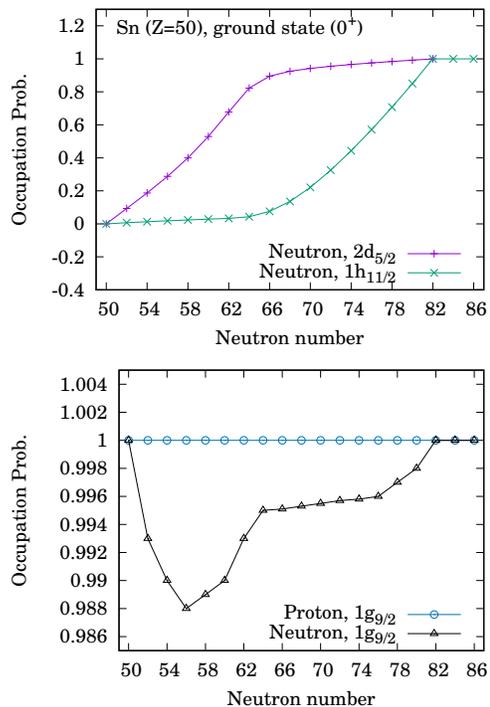}
\caption{The occupation probabilities of $\pi(1g_{9/2})$, $\nu(1g_{9/2})$, $\nu(2d_{5/2})$, and $\nu(1h_{11/2})$ orbits in our RHB-GS solutions for Sn isotopes.}
\label{fig:popy}
\end{figure}
\section{Results}
Before the RQRPA calculations for the M1 excitations in Sn isotopes, we briefly mention the single-particle occupation probabilities in the ground states, in order to check the effect of pairing correlations on the shell structure.
In Fig. \ref{fig:popy}, the occupation probabilities of our RHB solutions are summarized.
One can find that, for the neutron component, there are finite contributions of $\nu(2d_{5/2})$ and/or $\nu(1h_{11/2})$ orbits in some Sn isotopes.
This indicates the diffuseness of the Fermi surface at $N=50$ up to the $\nu(1g_{9/2})$ orbit.
Therefore, the neutron's M1 transitions are expected to appear in the 
$\nu(1g_{9/2} \longrightarrow 1g_{7/2})$,
$\nu(2d_{5/2} \longrightarrow 2d_{3/2})$, and/or 
$\nu(1h_{11/2} \longrightarrow 1h_{9/2})$ channels until the higher spin-partner orbits are occupied to block the M1 transition. 
Note also that the $\nu(2s_{1/2})$ is not active for M1 transition. 
On the other hand, for the proton component, the closed $Z=50$ shell at the $\pi(1g_{9/2})$ orbit keeps unbroken in the RHB calculation.
Thus, one can infer the proton's M1 transition in the $\pi(1g_{9/2} \longrightarrow 1g_{7/2})$ channel.
For comparison, in the recent shell-model study~\cite{Togashi2018}, which suggests the ground-state deformation in the light Sn isotopes,
finite holes in the $\pi(1g_{9/2})$ orbit are concluded, as well as a larger diffuseness of the neutron's Fermi surface is obtained.

\begin{figure}[t]
\includegraphics[scale=0.57]{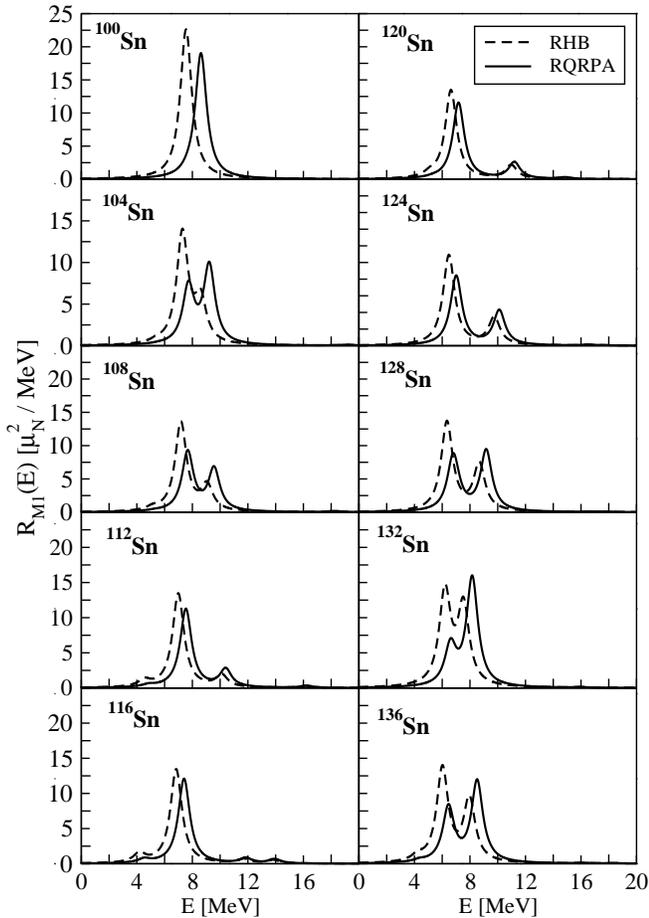}
\caption{The M1 transition strength function $R_{\rm M1}(E)$ for Sn isotopes, including the full RQRPA 
and the unperturbed (RHB) responses.
Calculations are based on the DD-PC1 parametrization of the RNEDF and  Gogny-D1S force
for the pairing correlations.}
\label{fig:Sn0B}
\end{figure}

In the following we present the results of the analysis of isotopic dependence of M1 excitation properties
along the Sn isotope chain.
Figure \ref{fig:Sn0B} shows the transition strength distribution function $R_{\rm M1}(E)$ for even-even $^{100-140}$Sn isotopes.
For comparison, in addition to the RQRPA results, the unperturbed RHB response without the contribution of the residual QRPA
interaction is shown. Thus one can observe a clear effect of the QRPA residual interaction which systematically shifts the M1
response toward higher excitation energies and the total $B_{\rm M1}$ strength somewhat reduces \cite{2019OP, 2020Oishi, 2020Kruzic}. 

As shown in Fig. \ref{fig:Sn0B}, for the $^{100}$Sn nucleus, the M1 response is represented by a single peak. This structure can be explained from two ingredients.
First, both for protons and neutrons, only the $\pi(1g_{9/2})$ and $\nu(1g_{9/2})$ orbits can be available for the M1 excitation.
The other orbits in view of the SO-splitting are fully occupied, and thus,
the one-body M1 operator cannot invoke their transitions.
Second, for the particle number 50, the pairing correlation vanishes, and thus, the M1-excitation energy is determined mainly by the SO-splitting energy \cite{2020Oishi}.
This SO splitting between $(1g_{9/2})$-$(1g_{7/2})$ is approximately common in the proton and neutron components. Thus, their M1-excitation energies coincide, and only a single peak is seen.

When the number of neutrons increases along the isotope chain $^{104-116}$Sn, the M1 response exhibits the second, higher-energy peak around $10$ MeV, as shown in Fig.~\ref{fig:Sn0B}.
This second peak is understood by considering the open (closed) shell of neutrons (protons).
Since the proton number Z=50 corresponds to the magic number, its ground state is characterized by the shell closure in the GS, and thus the proton M1 response does not change along the Sn isotope chain. We have verified that the lower peak is indeed attributed to this proton's excitation.
For neutrons in $^{104-116}$Sn, a more complicated M1 transition occurs due to the open neutron shell.
In addition to the IV-PV residual interaction, the pairing correlations become active,
resulting in noticeable shift of the  M1-excitation energy; also see Refs. \cite{2020Oishi, 2020Kruzic}.
Therefore, the neutron M1 transition deviates in energy from the proton one and generates the higher peak in the transition strength distribution.
From the analysis of relevant two-quasiparticle ($2qp$) components, we checked that this neutron-M1 excitation in $^{104-116}$Sn is mainly attributed to 
the $\nu(1g_{9/2} \rightarrow 1g_{7/2})$ transition. Note that, because of the diffuseness of Fermi surface due to the pairing interaction, a small amount of the $\nu(2d_{5/2} \rightarrow 2d_{3/2})$ transition also cooperates in the excitation.
Then, for the $^{116}$Sn nucleus, the neutron-M1 excitation is suppressed toward the minimum, because 
the $\nu(1g_{7/2})$ orbit is occupied in its GS: only the $\nu(2d_{5/2} \rightarrow 2d_{3/2})$ transition can contribute, in addition to the proton's one. 
Note that the $\nu(3s_{1/2})$ orbit also contributes in the GS solutions. 
However, this orbit is not active for the M1 transition.

From the $^{120}$Sn nucleus toward heavier Sn isotopes, the novel peak appears at $8-12$ MeV.
We emphasize that this higher peak is dominated by different transition, namely the $\nu(1h_{11/2} \rightarrow 1h_{9/2})$.
Up to the $^{132}$Sn nucleus, where the $\nu(1h_{11/2})$ orbit is filled in the GS, the corresponding M1 excitation strength grows up.
Then, for heavier systems, because the $\nu(1h_{9/2})$ orbit starts to be filled to block the $\nu(1h_{11/2} \rightarrow 1h_{9/2})$ transition, the transition amplitude becomes suppressed.
\begin{figure}[t]
\includegraphics[scale=0.33]{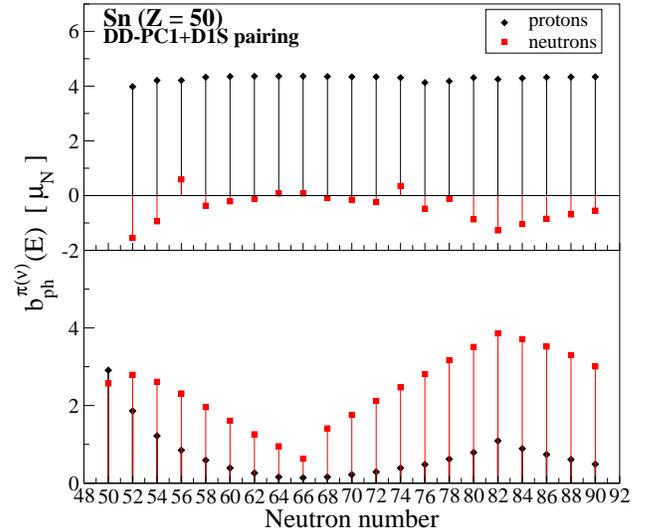}
\caption{The partial M1 transition strengths $b^{\pi(\nu)}_{ph}(E)$ for protons ($\pi$) and neutrons ($\nu$) of even-even $^{100-140}$Sn isotopes.
Partial strengths are evaluated at lower $E_<$ (upper panel) and $E_>$ higher energy peak (lower panel)  of double-peaked response function $R_{\rm M1}(E)$.}
\label{fig:Sn02B}
\end{figure}

To illustrate in more details the contributions from the relevant two-quasiparticle
configurations $(2qp)$ in the transition strength, in Fig. \ref{fig:Sn02B} we display the M1-transition amplitudes separately for the main neutron and proton transitions. 
Namely, the M1 transition strength is given by
\begin{equation}
  B_{\rm M1}(E_i) = \big\vert  b^{\pi}_{2qp}(E_i) + b^{\nu}_{2qp}(E_i) \big\vert^{2},
\end{equation}
where $E_i$ is the $i$th excitation energy obtained from the RQRPA. 
In the closed shell limit, the $2qp$ configuration reduces to the particle-hole ($ph$) configuration.
In Fig. \ref{fig:Sn02B}, partial contributions to the M1 strength $b^{\pi}_{2qp}(E_i)$ and $b^{\nu}_{2qp}(E_i)$ are plotted for the two major peaks in each Sn isotope, in 
support to the previous discussion on the relevant transitions. As one can see, the first
state is systematically dominated by the proton transitions ($\pi 1g_{9/2} \rightarrow 1g_{7/2}$), while neutron ones are very small. As discussed above, the second state
displays more subtle interplay between the neutron and proton transitions depending 
on the shell effects and occupation probabilities of relevant states, where in
most of the Sn isotopes the neutron transitions dominate.

\begin{figure}[t]
\includegraphics[scale=0.33]{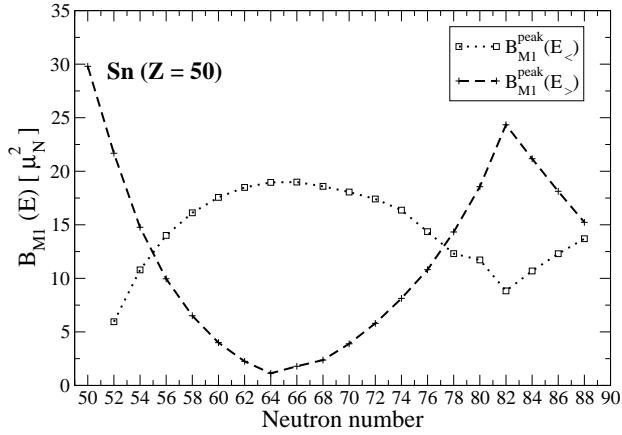}
\caption{The M1 transition strengths $B_{\rm M1}(E)$ for the two main peaks in even-even $^{100-140}$Sn isotopes, 
calculated using the R(Q)RPA with the DD-PC1 functional. The $E_<$ and $E_>$ energies correspond to the low- and high-energy dominant peaks in the response function $R_{\rm M1}(E)$.}
\label{fig:Sn02}
\end{figure}

In Fig. \ref{fig:Sn02} the transition strengths $B_{\rm M1}(E)$ are shown for two dominant peaks, denoted with $E_<$ and $E_>$ for lower and higher energy peaks, respectively. One can observe that the strengths of the two peaks display opposite trend along the isotope chain. Starting from $^{102}$Sn, the strength of the low- (high-) energy peak is increasing (decreasing), until its maximal (minimal) value obtained at $^{116}$Sn, and it continues decreasing (increasing) until doubly magic $^{132}$Sn, where the trend reverses again. This behavior can be understood from the analysis of the partial M1 transition strengths shown in Fig. \ref{fig:Sn02B}. For the low-energy state, the evolution of the $B_{\rm M1}$ strength with energy is a result of the destructive interference
between dominating nearly constant proton contributions and smaller neutron contributions that display variation along the isotope chain. On the other side, high-energy state evolution results from dominating neutron contributions followed by the same behavior of smaller proton contributions. More details about specific proton and neutron configurations involved are given in the discussion above. Clearly, rather different structures of the low- and high-energy state of the M1 transition strength result in different evolution of their properties along the Sn isotope chain.

\begin{figure}[t]
\includegraphics[scale=0.33]{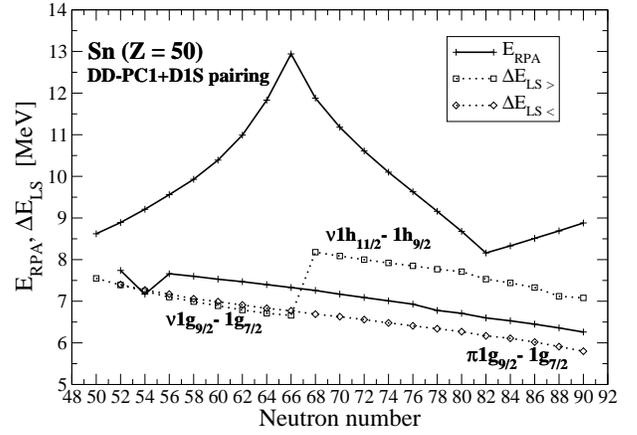}
\caption{The M1-excitation energies of $^{100-140}$Sn isotopes and the corresponding SO-splitting energies $\Delta E_{LS}$. The calculations
are based on the RHB plus RQRPA using DD-PC1 parameterization and D1S pairing force. The respective (nlj) quantum numbers are denoted for the SO-partner levels.
$\Delta E_{LS_<}$ and $\Delta E_{LS_>}$ represent the SO splitting contributions at lower and higher RPA energies, respectively.}
\label{fig:Sn03}
\end{figure}

Since the M1 excitations involve transitions between the spin-orbit (SO) partner states,
it is interesting to verify the relation between the SO energy splittings and R(Q)RPA excitation
energies of the M1 states.
Figure \ref{fig:Sn03} shows a comparison between the R(Q)RPA excitation energies with the spin-orbit splitting energies along the Sn isotope chain, calculated by,
\begin{equation}
\begin{aligned}
\Delta E_{LS} = E_{n\ell j_{<}} -  E_{n\ell j_{>}}, \label{eq:LSPOY}
\end{aligned}
\end{equation}
where $j_< = \ell - 1/2$ and $j_> = \ell + 1/2$ are spin-orbit partners of the dominant spin-flip ($j_<  \leftrightarrow  j_>$) transitions.
As we already know from previous studies \cite{2020Kruzic, 2020Oishi}, the M1 excitations induced by
spin-flip transitions hold $\ell_f = \ell_i$ orbital quantum numbers unchanged. As one can see in the figure,
the lower M1 state is very close to the proton SO splitting $(1g_{9/2})$-$(1g_{7/2})$, and the residual
R(Q)RPA interaction only slightly reduces the M1 excitation energies with respect to the relevant SO splittings. In the case of second M1 state, the situation is rather different. The respective transitions are dominated by $\nu(1g_{9/2} - 1g_{7/2})$ configuration until $^{116}$Sn and $\nu(1h_{11/2} - 1h_{9/2})$ configuration in heavier Sn isotopes. However, their R(Q)RPA-excitation energies are rather different than the SO splitting energies of neutrons. This result demonstrates the important role of the R(Q)RPA residual interactions, namely the IV-PV and pairing ones, which further increase the M1 excitation energies.
We can establish the rule for the isotope chain when the difference $\vert E_{R(Q)RPA} - \Delta E_{LS} \vert$ is minimal,
as a function of neutron number, as in case of $^{132}$Sn (N = 82), then the corresponding value of strength transition is maximal, as it can be seen in Fig. \ref{fig:Sn02B}. The opposite conclusion applies for the $^{116}$Sn (N = 66) nucleus, where the minimum of $B_{M1}$ value is obtained.

\begin{figure}[t]
\includegraphics[scale=0.33]{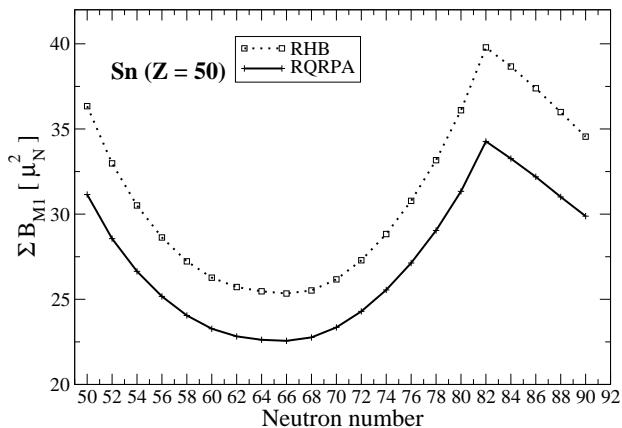}
\caption{The total M1 transition strengths for even-even nuclei in $^{100-140}$Sn isotope chain. 
The result is obtained with the R(Q)RPA using the DD-PC1 functional. The unperturbed RHB 
response is also shown.}
\label{fig:Sn01}
\end{figure}

An important benchmark test to validate our understanding of M1 transitions is comparison of the
model calculations with the respective experimental data. Although some data on M1 excitations in 
Sn isotopes are available from study in Ref. \cite{1998Govaert}, recent investigation
based on inelastic proton scattering provided more complete data on E1 and M1 strength
distributions along the even-even $^{112-124}$Sn isotope chain \cite{Bassauer2020}.
For comparison with the experimental data, in Fig. \ref{fig:Sn01}, the total M1 transition strength
is shown for Sn isotopes, 
$\sum_{i} B_{\rm M1}(E_{i}) = \int R_{\rm M1}(E) dE$, 
in the unit of $\mu^2_{\rm N}$.
The calculated strength $\sum B_{\rm M1}$ amounts $\approx 22$-$34~\mu^2_{\rm N}$
throughout the Sn isochain.
As already discussed before in detailed analysis of the main components in M1 transitions shown
in Fig. \ref{fig:Sn02B}, one can observe in the total M1 strength similar dependence on the neutron 
number, showing its minimum (maximum) for the $^{116}$Sn ($^{132}$Sn) nucleus. For comparison, both the results with and without residual QRPA interaction are shown, indicating the quantitative contribution of the residual interaction to the overall M1 transition strength amounting $\approx 5 \mu^2_{\rm N}$.

In Table \ref{table:Sn_TH_RPA}  we compare the calculated M1 excitation energies and transition strengths
for $^{116,120,124}$Sn with some data available from the study with polarized tagged photons \cite{1989Alarcon}
and nuclear resonance fluorescence (NRF) with polarized photons \cite{1998Govaert}. For  $^{116}$Sn, the RQRPA
strength strength distribution is dominated by a strong M1 peak at 7.33 MeV, that is very close to the measured
excitation at 7.92 MeV, though with considerably smaller B(M1) value. For $^{120}$Sn and $^{124}$Sn, 
double-peaked response function is obtained, resulting in reasonable agreement with available
experimental data (Table \ref{table:Sn_TH_RPA}). However, the B(M1) value for the low-energy
state for $^{120}$Sn appears considerably larger than experimental one. We note that in our model calculations
no quenching has been included in proton and neutron spin $g$ factors, suggested by some previous
studies \cite{1989Alarcon}. As pointed out in Ref.  \cite{1998Govaert},  lack of M1 transitions in NRF 
measurement is indication of a considerable fragmentation of the M1 strength over a large number of 
relatively weak individual transitions which could not be detected within the sensitivity of
experiments. 
\begin{table}[b]
\begin{center}
  \catcode`? = \active \def?{\phantom{0}} 
  \caption{Partial M1-excitation energy for selected states $E^{th.}_{peak}$ (in MeV) and the corresponding 
  transition strength $B^{th.}_{\rm M1}$   (in $\mu^{2}_{N}$) for
  $^{116}$Sn, $^{120}$Sn and $^{124}$Sn, calculated using the RQRPA (DD-PC1).   
  Experimental data for $^{116}$Sn and $^{124}$Sn are adopted from Ref. \cite{1998Govaert} while for $^{120}$Sn  the 
  data are from Ref. \cite{1989Alarcon}. } \label{table:Sn_TH_RPA}
  \begingroup \renewcommand{\arraystretch}{1.7}
  \begin{tabular*}{\hsize} { @{\extracolsep{\fill}} lcccc } \hline \hline
           &  $E^{th.}_{peak}$    &  $B^{th.}_{\rm M1}$   &$E^{exp.}_{peak}$     &$B^{exp.}_{\rm M1}$    \\ \hline
 ~$^{116}$Sn    &    $?7.33$                       &$18.99$                                  &              7.92                   &          0.28                              \\
~                         &$12.94$                           &$?1.79$                                  &          *                        &             *                                 \\ \hline
 ~$^{120}$Sn    &$?7.17$                           &$18.06$                                  &         7.3 - 9.3              &           8.8                           \\
 ~                        & $11.18$                          &$?3.89$                                  &             *                       &            *                                  \\ \hline
  ~$^{124}$Sn   &  $?7.01$                         &$16.70$                                  &           6.80                     &           *                                         \\
  ~                       &  $10.10$                        &$?8.13$                                   &           8.25                     &           *                                     \\ 
                            &                                       &    &   &  $\sum B^{exp.}_{\rm M1} = 0.61 $   \\ 
 \hline \hline
  \end{tabular*}
  \endgroup
  \catcode`? = 12 
\end{center} 
      \small{*}
      No data provided per individual resonant peak.
\end{table}

In the new experimental study with inelastic proton scattering, the M1 transition data
are provided for $^{112,114,116,118,120,124}$Sn \cite{Bassauer2020}. Although the measurements
provide rather broad strength distributions, for $^{120,124}$Sn double-hump structures
can be observed, that are in qualitative agreement with the RQRPA results. 
For $^{112,114,116,118}$Sn, the experiment results
in broad structure with pronounced maximum, consistent with a single-dominant M1 peak 
in the RQRPA calculation. For a more quantitative comparison, Table \ref{table:Sn_exp_Bassauer}
shows the total RQRPA transition strength for M1 excitations in $^{112-124}$Sn
(also see Fig. \ref{fig:Sn01}), together with the experimental data from inelastic proton
scattering \cite{Bassauer2020}. We conclude that the calculated total B(M1) strength appears 
larger than the experimental values, and no systematic dependence on the neutron 
number, as obtained in the RQRPA calculations, is obtained from the experiment.
This comparison indicates that some of the M1 strength may be missing from the experiment. 
In fact, as explained in Ref. \cite{Bassauer2020},
the experimental results above the neutron separation energies have limited accuracy because of
the similarity of the M1 and the phenomenological continuum angular distributions in the multipole 
decomposition analysis (MDA).

The present results on M1 transitions, together with new experimental
data \cite{Bassauer2020} represent considerable progress  compared to previous studies of the M1-quenching effect,
because the discrepancy between theory and experiment is significantly reduced, e.g., for $^{114}$Sn, the ratio 
of total B(M1) values is only 1.15, for $^{118}$Sn it is 1.23, etc. (see Table \ref{table:Sn_exp_Bassauer}).
For example, previous systematic study based on Gogny interaction indicated
that factor of 2 is needed to reproduce the experimental data \cite{2016Goriely}. 
Table \ref{table:Sn_exp_Bassauer} also shows the ratio of the effective $g$ factor including quenching ($g_{eff}$) 
with respect to that for free nucleons ($g_{free}$), where the quenching factor is determined in order to
reproduce the experimental data for M1 transition strength in each even-even isotope $^{112-124}$Sn.
Assuming the general quenching applies to all gyromagnetic factors involved in the M1 transition operator,
the results show $g_{eff}/g_{free}$=0.80-0.93, that is 
higher than previously reported values $g_{eff}/g_{free} \approx$ 0.6-0.75 \cite{Heyde2010,1989Alarcon,2009Vesely,2010Nest,1998VNC,Ichimura2006}, i.e., smaller quenching of the $g_{free}$ factor is needed
to reproduce inelastic proton scattering data on M1 transitions. 
As discussed above, since part of the experimental data above the neutron threshold may be missing, the results of the present study
indicate that the actual quenching of the $g$ factors in nuclear medium may be very small or even 
negligible in comparison to the $g$ factors for the free nucleons. Clearly, additional experimental studies are required to confirm this.

We also performed a complementary analysis of the quenching of $g$ factors, by investigating the Gamow-Teller ($1^{+}$) transition 
$^{100}$Sn $\to$ $^{100}$In, recently investigated from the measurement of the half-life and decay energy for the decay of
$^{100}$Sn \cite{Hinke2012}. The resulting transition strength amounts $B(GT, exp.) =$ 9.1+2.6/-3.0. 
By employing the proton-neutron RQRPA from Ref.~\cite{Vale2020}, we obtain the respective strength $B(GT^+)=$13.67,
i.e., the quenching of 0.82 is needed in the $g$ factor to reproduce the experimental data. Therefore, the result obtained
using the GT transition from $^{100}$Sn  seems to be consistent with our analysis of M1 transitions in Sn isotopes.
\begin{table}[b]
\begin{center}
  \catcode`? = \active \def?{\phantom{0}} 
  \caption{The total RQRPA (DD-PC1) transition strength $\sum B^{th.}_{\rm M1}$ in $\mu^{2}_{N}$ units for $^{112-124}$Sn 
  in comparison to the experimental data from inelastic proton scattering \cite{Bassauer2020}. Last column ($g_{eff}/g_{free}$)
  shows the quenching of the $g$ factors of the free nucleons, needed to reproduce the experimental data on M1 transition strengths.}
  \label{table:Sn_exp_Bassauer}
  \begingroup \renewcommand{\arraystretch}{1.7}
  \begin{tabular*}{\hsize} { @{\extracolsep{\fill}} lccccc } \hline \hline
           &  $\sum B^{th.}_{\rm M1}$    &  $\sum B^{exp.}_{\rm M1}$  & $g_{eff}/g_{free}$  \\ \hline
 ~$^{112}$Sn    & $22.81$                       &$14.7(1.4)$    &  0.80 \\
 ~$^{114}$Sn    & $22.61$                       &$19.6(1.9)$    &  0.93\\
 ~$^{116}$Sn    & $22.56$                       &$15.6(1.3)$    &  0.83 \\
 ~$^{118}$Sn    & $22.76$                       &$18.4(2.4)$    &  0.89 \\
 ~$^{120}$Sn    & $23.34$                       &$15.4(1.4)$    &  0.81 \\
 ~$^{124}$Sn    & $25.55$                       &$19.1(1.7)$    &  0.86 \\            
 \hline \hline
  \end{tabular*}
  \endgroup
  \catcode`? = 12 
\end{center} 
 \end{table}

\begin{figure}[t]
\includegraphics[scale=1.0]{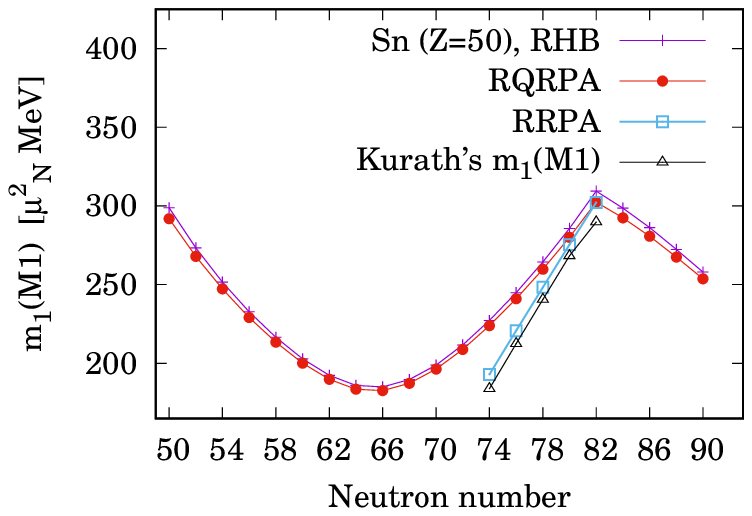}
\includegraphics[scale=1.0]{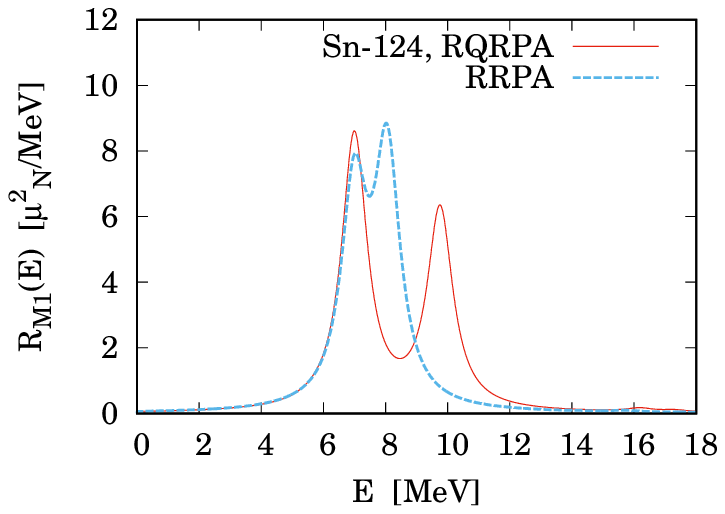}
\caption{(Top) Energy-weighted summation $m_{1}({\rm M1})$. For $^{124-132}$Sn, the corresponding results with Kurath's sum rule are also displayed.
(Bottom) M1 strength distribution of $^{124}$Sn with and without the pairing interaction.}
\label{fig:CIWYG}
\end{figure}

\section{\label{sec:SUM} Summary}
In this work we have investigated the evolution of M1 ($0^{+} \rightarrow 1^{+} $) excitations in even-even Sn isotopes based on the
RHB+R(Q)RPA formulated in the framework of relativistic nuclear energy density functional \cite{2020Kruzic, 2020Oishi}. 
The M1 excitations induced by $\hat{\bm{\mu}}_{1\nu}$ operator, mainly by its spin component, are governed by single 
particle spin-flip transitions $j_< \leftrightarrow j_>$ between corresponding spin-orbit partners.
Model calculations show that along the Sn isotope chain the M1 transition strength distribution
is characterized by the interplay between single- and double-peak structures. In the latter case, detailed analysis of
the $2qp$ components showed that the first peak is dominated by proton spin-flip transition, while the
second peak displays an interplay between dominating neutron spin-flip transition and smaller proton transition. 
The evolution of the M1 transition strength of the two main peaks is governed by the subtle effects 
of the single-particle structure, pairing correlations, respective occupation probabilities, and RQRPA 
residual interaction.
Comparison with available experimental data shows that independent neutron and proton spin-flip spectra are correctly identified, 
single and double-peaked distribution of response function $R_{\rm M1}(E)$  is reasonably well reproduced.
The calculated peak positions $E^{th.}_{peak}$ show 1-2 MeV discrepancy, 
that could be further fine-tuned through
additional adjustments of the strength parameter in the isovector-pseudovector channel, $\alpha_{IV-PV}$. 

While in the previous studies the calculated M1-transition strengths considerably overestimated the experimental values, comparison of the RQRPA results for the total $B_{\rm M1}$ strength with the new data on Sn isotopes from inelastic proton scattering \cite{Bassauer2020} shows that differences are smaller than previously understood. 
Our analysis showed that discrepancy between model calculations and experiment are considerably reduced, i.e., the quenching of the $g$ factors for the free nucleons needed to reproduce the experimental data amount $g_{eff}/g_{free}$=0.80-0.93.
Considering the fact that some of the $B_{\rm M1}$ strength above the neutron threshold may be missing in
the proton scattering experiment due to the reported limitations in accuracy \cite{Bassauer2020}, our analysis
provides an indication that future experimental 
studies could confirm that actually very small or even no modifications of the $g_{free}$ factor are needed when
applied in the nuclear medium in finite nuclei. Therefore, we sincerely hope that our present work can serve as guidance and
motivation for the experimental community to systematically explore M1 resonant excitations, and in particular
to reduce the uncertainties above the neutron threshold. Finally, complete understanding of M1 transitions within 
the framework used in this study, will also allow systematic large-scale calculations for the radiative neutron
capture cross sections of relevance for the nucleosynthesis.

\begin{acknowledgments}
This work is supported by 
the ``QuantiXLie Centre of Excellence'', a project co-financed by 
the Croatian Government and European Union through 
the European Regional Development Fund, the Competitiveness and Cohesion 
Operational Programme (KK.01.1.1.01). 
\end{acknowledgments}

\appendix*
\section{Energy-weighted summation of the M1 strength}
As a verification of the present model calculations, we compare the 
energy-weighted summation (EWS) of the M1 transition strength with the 
Kurath M1-sum rule given in Ref. \cite{1963Kurath}. The EWS reads
\begin{equation}
m_{1}({\rm M1}) = \int E \cdot R_{\rm M1}(E) dE.
\end{equation}
In Fig. \ref{fig:CIWYG}, our results of the EWS are summarized.
One can find that the RHB and RQRPA results show a similar behaviour, that is consistent to the non-energy-weighted
summation $\sum B_{\rm M1}$ shown in Fig. \ref{fig:Sn01}. 
On the other side, the Kurath M1-sum rule \cite{1963Kurath}, is given by
\begin{eqnarray}
&& m_{1}({\rm M1,~Kurath}) \nonumber  \\
&& \cong \frac{3\mu^2_{\rm N}}{4\pi} \left( g^{\rm IV}_s + g^{\rm IV}_l \right)^2 \sum_{i} (-a_{\rm SO}) \Braket{{\bf l}(i) \cdot {\bf s}(i)}, \label{eq:Kurath}
\end{eqnarray}
where $g^{\rm IV}_s=-4.706$, $g^{\rm IV}_l=1/2$, the bracket means the expectation value for the ground state, and the summation $\sum_{i}$ includes only the M1-active protons and neutrons. The $a_{\rm SO}<0$ within the unit of MeV indicates the simple spin-orbit splitting parameter utilized by Kurath \cite{1963Kurath}, that is, $V_{\rm SO}(i) \equiv a_{\rm SO}{\bf l}(i) \cdot {\bf s}(i)$ for the $i$th nucleon. 
According to Eq. (\ref{eq:Kurath}), when the M1-active nucleons are in the single orbit with the same $\Braket{{\bf l}(i) \cdot {\bf s}(i)}$ value, the $m_{1}({\rm M1})$ value is simply proportional to the number of those nucleons. 

For a comparison of our results with the Kurath's sum rule, one point is notified.
The original Kurath sum rule refers to the isovector-M1 transitions.
On the other side, our summations include both proton and neutron transitions,
and thus, the isoscalar-M1 contributions are also included.
However, we confirmed that this isoscalar-M1 contribution exhausts only
$\cong 1$ \% of the total M1 summations, and thus, it is safely neglected
within the present comparison. More details about the dominance of
isovector-M1 mode are also available in Ref. \cite{2020Kruzic}.

Using the Kurath's ansatz, the most suitable example for verification of our model calculations is a set of nuclei $^{124-132}$Sn. From our RQRPA calculation, their M1 excitations can be mostly attributed to the $\nu(1h_{11/2}\longrightarrow 1h_{9/2})$ and $\pi(1g_{9/2} \longrightarrow 1g_{7/2})$ transitions.
Because both the proton and neutron components are active, the summation in Eq. (\ref{eq:Kurath}) is represented as,
\begin{eqnarray}
\sum_{i}(-a_{\rm SO}) \Braket{{\bf l}(i) \cdot {\bf s}(i)}
&=& \sum_{i\in \pi}(-a^{\pi}_{\rm SO}) \Braket{{\bf l}(i) \cdot {\bf s}(i)} \nonumber  \\
&+& \sum_{i\in \nu}(-a^{\nu}_{\rm SO}) \Braket{{\bf l}(i) \cdot {\bf s}(i)}, \label{eq:VSTEWY}
\end{eqnarray}
for $\nu(1h_{11/2})$ and $\pi(1g_{9/2})$ orbits. 
In Fig. \ref{fig:CIWYG}, we plot the Kurath's sum rule, for which the parameter $a_{\rm SO}$ is fixed to reproduce the spin-orbit splitting, $\Delta E_{LS}$ in Eq. (\ref{eq:LSPOY}), obtained in the RHB ground states of $^{124-132}$Sn. 
By comparing our calculation with Kurath's sum rule, there is a finite gap: the RQRPA and RHB results overshoot the Kurath's $m_1({\rm M1})$ values in Fig. \ref{fig:CIWYG}. 
This gap can be attributed to the two ingredients, namely (i) the transitions to the higher, or even continuum states, and (ii) the pairing correlation. 
For the point (i), there can be a small but finite contribution of, e.g. $\nu(1h_{11/2}\longrightarrow Ih_{9/2})$ where $I>1$ in the RQRPA calculations. These components are neglected in Eq. (\ref{eq:VSTEWY}). 
For the (ii) pairing correlation, one should notice that the M1-active neutrons and protons may have the mixture of orbits, i.e. the smearing of the Fermi surface. This situation is not considered in Kurath's ansatz \cite{1963Kurath}, in which the M1 transition only from the single orbit is considered. 

In order to check the pairing effect on the EWS values, for $^{124-132}$Sn, we repeated the same calculations but completely neglecting the pairing interaction both in the RHB and RQRPA, leading to the Hartree+RRPA calculation. The respective RRPA results, shown
in Fig. \ref{fig:CIWYG}, now result in a good agreement with the Kurath sum rule.
Therefore, except the small gap due to the transitions to the higher states, 
our calculation reproduces the Kurath's sum rule.

By comparing the RQRPA and RRPA results in Fig. \ref{fig:CIWYG}, one can find that the pairing interaction enhances the EWS values, $m_1({\rm M1})$. 
This enhancement is due to the energy shift. 
In the bottom panel in Fig. \ref{fig:CIWYG}, for example, we plot the M1-strength distribution, $R_{\rm M1}(E)$, for $^{124}$Sn. 
There, it is shown that the pairing interaction increases the M1-excitation energy of the second peak: this effect has also been shown in our previous studies \cite{2020Kruzic, 2020Oishi}. 
We note that the EWS value results from a competition of (i) the energy shift and (ii) the change of M1 strength, both invoked by the pairing interaction. 
Therefore, the $m_1({\rm M1})$ value can also be increased for the Sn nuclei, where the reduction of $B_{\rm M1}$ by the pairing is minor compared with the energy shift. 
Note that, however, this conclusion is not common for the other systems. 
Indeed, in the Ca isochain in Ref. \cite{2020Oishi}, the opposite conclusion on the EWS was obtained, because the M1-energy shift by the pairing interaction is not sufficiently large to increase the $m_1({\rm M1})$ values. 

\bibliographystyle{apsrev4-1}
\bibliography{main}

\end{document}